\documentclass[a4paper]{article}
\usepackage{RR}
\usepackage{hyperref}
\usepackage{makeidx,graphicx}

\RRdate{Octobre 2006} 
\RRauthor{
Yongho Seok \and Thierry Turletti} 
\authorhead{Seok \& Turletti}
\RRtitle{Mécanismes de Transmission Multipoint\\
pour Réseaux Locaux Sans Fil IEEE 802.11\\}
\RRetitle{Practical Rate-Adaptive Multicast Schemes for Multimedia
over IEEE 802.11 WLANs} 
\titlehead{Practical Multicast Schemes for IEEE 802.11}
\RRresume{Le standard IEEE 802.11 est inefficace pour la
transmission multimédia en multipoint. En particulier, les paquets
multipoints sont envoyés en boucle ouverte de la même manière que
les paquets broadcast. L'absence d'acquittements rend impossible la
mise en \oe uvre de mécanismes de contrôle de congestion, de
mécanisme de fiabilisation de la transmission ainsi que
d'algorithmes d'adaptation du débit de transmission physique. Dans
ce rapport, nous proposons de nouveaux mécanismes de transmission
multipoint qui se basent sur une approche {\em leader} pour renvoyer
des acquittements. Nous nous interessons à des solutions pratiques
qui sont suceptibles d'être implantés dans les cartes réseaux sans
fil actuelles et futures et qui restent compatibles avec les
stations IEEE 802.11 standards. Nous proposons deux mécanismes pour
adapter le débit de transmission physique des flots multipoints: un
mécanisme simplifié appelé LB-ARF et un mécanisme plus robuste
appelé RRAM. Nos simulations montrent que pour des environnements
statiques, un mécanisme aussi simple que LB-ARF suffit pour obtenir
de bonnes performances. Le mécanisme RRAM est quant à lui aussi
efficace dans des environnements statiques que lorsque les stations
sont mobiles.
 }
 \RRmotcle{IEEE 802.11, algorithmes d'adaptation du débit de transmission physique, transmission multipoint}

\RRabstract{The current IEEE 802.11 standard does not address the
basic requirements of multicast communication. More specifically,
multicast packets are sent in an open-loop fashion as broadcast
packets, i.e., without any acknowledgements. This basic multicast
transmission mechanism prevents the implementation of {\em
congestion control}, {\em transmission reliability}, and {\em
physical data rate adaptation} algorithms. In this paper, we propose
new mechanisms based on the {\em leader based} approach to enhance
the legacy multicast transmission scheme in WLANs. We focus on
practical solutions that can be deployed in current and future WiFi
devices and are compatible with legacy 802.11 devices. We propose
two mechanisms to adapt the PHY data rate of multicast flows: the
{\em simplest leader-based mechanism (LB-ARF)} and the {\em Robust
Rate Adaptive Multicast mechanism (RRAM)}. Our simulations show that
for static environments, LB-ARF and RRAM can achieve high multicast
throughput and fairness between the set of multicast receivers.
LB-ARF is sufficient to outperform the legacy multicast mechanism
when the stations are fixed. RRAM improves the reliability of the
multicast transmission and obtains high throughput independently of
the number of multicast receivers and the maximal speed of
stations.} \RRkeyword{IEEE 802.11, leader-based approach, multicast
transmission, PHY rate adaptation} 
\RRprojet{Planète} 
\RRtheme{\THCom} 
\URSophia 
\begin{document}
\makeRR   

\section{Introduction}
\label{sec:introduction}

The IEEE 802.11 protocol suite aka WiFi is very popular today
because it represents a cost effective solution to provide
relatively high bandwidth connectivity to wireless LANs (WLANs).
Most of today's current personal digital assistants (PDAs) and
laptops have by default a WiFi interface. Moore's law and advances
in multimedia communication techniques (e.g., compression) make
these devices increasingly more capable of handling live multimedia
applications. As hot spots become more ubiquitous, people on the
move will be able to use their wireless devices (PDAs, cell phones,
etc) to receive multimedia data (e.g., watch live broadcasts of
news, presentations, etc).

It is well known that multicasting flows instead of streaming them
individually results in a much more efficient use of the shared
wireless medium. Whereas all these new applications are very likely
to appear soon with upcoming WiMAX or DVB-H enabled devices, the
IEEE 802.11 standard does not address multicast data
requirements~\cite{diego}. In particular, the current 802.11
standard sends multicast packets similarly to broadcast packets,
i.e., without acknowledgements. This basic multicast transmission
mechanism poses three main problems, which are described below.

The most critical one is the fact that, without any feedback
mechanism, {\em congestion control} is not possible for multicast
flows resulting in unfairness with other concurrent unicast flows.
In IEEE 802.11, unicast flows use the DCF access scheme, where
contention windows (CW) change dynamically to adapt to the
contention level: Upon each collision, a node doubles its CW to
reduce further collision risks. Upon a successful transmission, the
CW is reset, assuming that the contention level has dropped. Without
feedback, multicast flows are not able to adapt their contention
window according to the network state. Consequently, they can not
only starve concurrent unicast flows but also severely congest the
network.

The second problem concerns {\em transmission reliability}. Although
multimedia applications can tolerate a certain percentage of packet
loss, their performance may degrade severely in the presence of
persistent transmission errors or high channel load. Indeed,
contrary to unicast transmissions, no MAC retransmission mechanism
is provided for multicast. Corrupted frames (due to transmission
errors or collisions) are simply dropped.

The third problem deals with {\em physical data rate selection}. To
achieve high performance under varying channel conditions, IEEE
802.11 devices adapt their PHY transmission rate dynamically.
Several mechanisms have been proposed in the literature such as
RBAR~\cite{rbar} or CLARA~\cite{clara} or the commercial
ARF~\cite{arf} protocol. But these mechanisms are not usable with
the current open-loop multicast transmission protocol. Indeed, most
commercial access points (APs) to-date use a fixed and relatively
very low transmission rate for multicast transmissions. Such a case
exhibits the 802.11 anomaly~\cite{anomaly}, and the performance of
other unicast stations is seriously degraded as the multicast
traffic overwhelms the wireless bandwidth due to the fixed and low
data rate~\cite{diego}.

As discussed in the following section, several solutions have been
proposed recently to solve these problems, but none of them can
actually be used today because of implementation issues or
compatibility problems with legacy 802.11 devices. In this paper, we
focus on practical solutions that can be adopted by current and
future 802.11 devices.

The remainder of this paper is organized as follows. In Section
\ref{sec:related-works}, we present a review of solutions proposed
so far to enhance the 802.11 multicast transmission mechanism and we
discuss implementation issues of solutions. Section III and Section
IV describe our solutions composed of a leader election protocol and
two multicast PHY rate adaptation algorithms. Especially, in
Section~\ref{sec:LB-ARF}, we describe the simplest solution that
could be used with current devices in static environments. Then we
describe in Section~\ref{sec:rram} the proposed Robust Rate Adaptive
Multicast mechanism (RRAM) which aims to provide an efficient
solution for mobile 802.11 environments and dynamic channel
conditions. In Section~\ref{sec:performance}, we evaluate the
performance of RRAM against the current IEEE 802.11 multicast
transmission protocol. Finally, we conclude in
Section~\ref{sec:conclusion} and present directions for future work.

\section{Related work and Discussion}
\label{sec:related-works}

One of the alternatives to improve the current 802.11 multicast
mechanism for reliable transmissions is the leader-based reliable
multicast scheme~\cite{lbp}. In a nutshell, this solution proposes
to select one of the receivers to send acknowledgement frames back
to the sender. To transmit a multicast frame, the AP first sends a
non-standard multicast-RTS frame. If a leader is ready to receive
the multicast frame, it replies with a CTS frame. Other stations
send a non-standard NCTS (Not Clear to Send) frame if they are not
ready to receive the multicast frame. In other cases, the leader and
other stations do not send any frame. If the AP hears a CTS from the
leader, it starts a multicast transmission. Else, it performs a
backoff to retransmit the multicast frame. Upon receiving the
multicast frame, if the leader receives it without error, it sends
an ACK frame. Otherwise, the leader and other stations send a
non-standard negative acknowledgment (NAK) frame. As with regular
unicast transmissions, the multicast sender can use a PHY rate
selection mechanism such as ARF~\cite{arf} and lost frames can be
retransmitted as it is the case for unicast flows. Furthermore, the
leader-based approach provides fairness with other concurrent
unicast flows because the same algorithm also adjusts the contention
window according to the perceived congestion conditions.

Another approach to solving the problem of lack of congestion
control, proposed by Choi et al, dynamically adapts the contention
window for multicast frames according to the number of competing
stations in the wireless LAN~\cite{ccnc}. However, this solution
does not improve transmission reliability and still uses a fixed PHY
data rate.

In \cite{jose}, Villalon et al. have proposed a solution, called
auto rate selection mechanism (ARSM), to solve the three problems
identified for the IEEE 802.11 multicast mechanism. Basically, ARSM
dynamically selects the multicast PHY rate based on channel
conditions perceived by the receiving stations. In order to reduce
the rate of feedback collision, ARSM uses the SNR value of the
station to decide when the feedback frame is transmitted. The
station with the worst SNR has the highest priority to send its
feedback frame. Then, the AP can select the station with the lowest
SNR value as the leader. The main flaw of the ARSM mechanism is that
it uses new control frames for the feedback mechanism which makes it
incompatible with current 802.11 stations.

The easiest way to solve the three problems identified with the
current 802.11 standard's multicast mechanism is to emulate unicast
transmission using a leader-based approach, which, in a nutshell,
means that one of the receiving stations is responsible to send
acknowledgements on behalf of the intended receiving stations. This
feedback is used to trigger possible retransmissions, adapt the
contention window, and select the PHY data rate. Our possible leader
selection policy is to choose the receiver with the worst channel
conditions. However, the overhead associated with leader election
increases with the number of multicast receiving stations, so, it is
not efficient to choose a new leader for each new transmission. On
the other hand, the algorithm to select the PHY data rate requires
per-packet feedback. The fact that these algorithms run at two
different timescales can cause situations where the current leader
does not correspond to the receiver which experiments the worst
channel conditions. In particular, for a simple PHY rate selection
algorithm such as ARF~\cite{arf} used in combination with a
leader-based mechanism, when the leader decides to increment the PHY
rate, it is not guaranteed that other receivers can afford the rate
increment. In extreme cases, some receivers can even become
disconnected from the data session.  A way to prevent such a problem
is to allow feedback from any receivers before taking critical
decisions such as rate increase.

It is important to note that  the PHY data rate selection algorithm
supplements the leader selection mechanism because it enhances
transmission reliability even when the current leader does not
correspond to the worst receiver in the group. Two different types
of statistics can be used to select the PHY data rate:  statistics
on previous packets sent (used by ARF~\cite{arf}/AARF~\cite{aarf})
or SINR statistics (used by RBAR~\cite{rbar}, CLARA~\cite{clara}).
There are pros/cons for both approaches. The main problem with
ARF/AARF is that they are not as reactive than SINR-based solutions
and may generate bad experiments or periodic losses. On the other
hand, SINR-based solutions can be device-dependent, and the SINR
information  may sometimes be imprecise. The mechanisms proposed in
this paper consider both approaches.

It is also important to propose solutions that do not use negative
acknowledgement (NAK) frames like the leader-based reliable
multicast scheme~\cite{lbp} because they have important
implementation issues. In particular, the decision to send NAK
frames has to be immediate (and sometimes wired in the hardware).

In this paper, we focus on pratical solutions that try to limit
implementation issues and keep compatible with legacy IEEE 802.11
devices. Leader-based mechanisms are composed of two main
algorithms, that select the leader and select the PHY data rate. We
first describe the leader election protocol in Section \ref{sec:lep}
which will be used by the multicast PHY rate adaptation mechanisms
in Section \ref{sec:rate}.

\section{Leader Election Protocol}
\label{sec:lep}

The proposed Leader Election Protocol (LEP) dynamically selects the
receiving station with the worst current channel conditions as the
leader. The LEP mechanism is based on IGMP and consists of the
following four phases.

\subsection{Collection Phase}

To select the leader, LEP needs to estimate the channel conditions
of each multicast receiver. To this end, multicast receivers
periodically send modified IGMP Membership Reports (MR) that include
the SINR indication (7 bits) within the MRT\footnote{MRT specifies
the maximum allowed time before sending a responding report but is
meaningful only in an IGMP Membership Query message sent by a
multicast router. In other messages, the MRT is set to 0 by a sender
and ignored by receivers.} field, see Figure \ref{fig:membership}.
The duplicated bit (D bit) reserved for the leader reelection is
reset to 0.

\begin{figure}

\begin{center}
\caption{Modified IGMP format (for MR and GSQ).}
\label{fig:membership}
\end{center}
\end{figure}

When the AP receives an IGMP Membership Report with a non-zero MRT,
it assumes that the station supports the LEP mechanism. Then, the AP
stores the multicast group address, the MAC address and the SINR of
the station.

\subsection{Election Phase}

Whenever receving an IGMP Membership Reports, the AP chooses the
station with the lowest SINR. If the current leader is not the worst
station, the AP sends a modified IGMP Group Specific Query (GSQ)
which includes the SINR of the selected worst station within MRT
field. If more than one stations have the same lowest SINR value,
the AP sets the duplicated bit to 1. Otherwise the duplicated bit is
set to 0. The duplicated bit corresponds to the {\bf D bit} of
Figure \ref{fig:membership}.

Once receving IGMP Group Specific Query, each mobile station checks
the source IP address of the IGMP Group Specific Query. If the
source IP address does not correspond to the AP, the packet is
considered as a legacy IGMP Group Specific Query coming from the
multicast routers. So, each multicast receiver sends the legacy IGMP
Membership Report after some delay time. Otherwise, each mobile
station carries out the following $Confirmation~Phase$.

\subsection{Confirmation Phase}

Through the MRT field of the IGMP Group Specific Query, each
multicast receiver can know the previous SINR of the new elected
leader. So, each multicast receiver compares the reported SINR
during the $Collection~Phase$ with the SINR of new elected leader.
Then, the multicast receiver having the same SINR does the
followings according to the duplicated bit.

\begin{itemize}
\item{ \bf Duplicated bit == 0} : send the additional IGMP Membership
report with same SINR to confirm the leader election. The duplicated
bit of the IGMP Membership Report is reset to 0.
\item{ \bf Duplicated bit == 1} : send the additional IGMP Membership
report with the random number instead of same SINR to carry out the
$Reelection~Phase$. Duplicated bit of the IGMP Membership Report is
set to 1.
\end{itemize}

Finally, if the AP receives the IGMP Membership Report with same
SINR value, then the AP terminates the leader election algorithm.

Else if the AP does not receive the IGMP Membership Report, the AP
will retransmit the IGMP Group Specific Query with the confirmation.
Otherwise, the AP will try the $Reelection~Phase$. This confirmation
phase is very important. Because the IGMP Group Specific Query is
simply broadcasted without any acknowledgement.

\subsection{Reelection Phase}

If the AP receives the IGMP Membership Report of which the
duplicated bit is set to 1, it does not change the SINR statistic
about this station because this IGMP Membership Report is just used
for the leader reelection. Then, the AP sends the additional IGMP
Group Specific Query. However, the MRT field of IGMP Group Specific
Query is equal to one of previously received IGMP Membership Report
(i.e., the random number choosed by station). The duplicated bit of
this Group Specific Query is set to 0.

\section{Multicast PHY Rate Adaptation Mechanism}
\label{sec:rate}

In this section, we propose two PHY rate adaptation mechanisms,
LB-ARF and RRAM both of which work in tandem with LEP.

\subsection{LB-ARF}
\label{sec:LB-ARF}

First we propose the simplest leader-based mechanism (or LB-ARF) for
rate-adaptive multicast. In LB-ARF, the leader elected by LEP sends
an acknowledgement frame to the AP once a multicast frame has been
successfully received.

Then, the AP controls the multicast PHY rate similarly to ARF. When
the timer expires or once 10 consecutive ACKs are received, the
multicast PHY transmission rate is increased to the next higher rate
and the timer is reset. When losses occur, after two consecutive
lost frames, the PHY transmission rate is decremented and the timer
is restarted.

If the worst station is always elected as the leader, LB-ARF can
provide throughput fairness between multicast receivers but also
achieves high multicast throughput. However, LB-ARF is not
appropriate for mobile environments. Because the channel conditions
of the receivers are changing quickly.

\subsection{RRAM}
\label{sec:rram}

The Robust rate adaptive multicast mechanism (RRAM) aims to extend
LB-ARF targeting dynamic environments (e.g., due to mobility).

As we mentioned in Section~\ref{sec:related-works}, it is important
to allow feedback from any receivers before taking critical
decisions such as rate increase because the leader may not have the
worst channel conditions at this time. So, RRAM requires a mechanism
to inform all receivers that the leader is about to increase the PHY
transmission rate.

To this end, RRAM requires a minor modification in the MAC header of
multicast data frames. The 802.11 MAC header contains a 16-bit field
called $Sequence~Control$ field which is composed of two subfields:
$Sequence~Number$ and $Fragment~Number$. These subfields are used
for retransmission and fragmentation only in point-to-point
transmissions.

\begin{figure}
\begin{center}
\includegraphics[width=9.0cm]{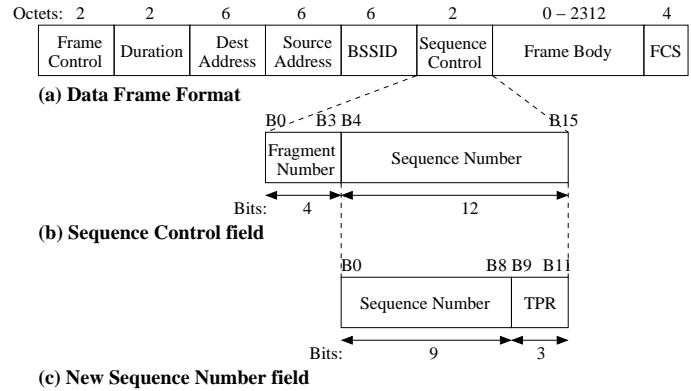}
\caption{New Sequence Number field in multicast frame.}
\label{fig:sequence}
\end{center}
\end{figure}

Therefore, in RRAM we propose to use the $Sequence~Control$ field to
inform multicast receivers that the leader is about to increase the
PHY transmission rate. As shown in Figure~\ref{fig:sequence}, a new
3-bit field termed $Target~Probe~Rate~(TPR)$ is added and the
$Sequence~Number$ field is encoded in 9 bits in the multicast case.
The $TPR$ field will encode the value of the PHY transmission rate
that the leader wants to enforce. Since the length of the $TPR$
field is 3 bits, 8 different PHY transmission rates can be
supported.

So, when multicast receivers receive data frames with a $TPR$ value
larger than current PHY transmission rate, they can deduce that the
leader requires a PHY rate increase. In this case, each multicast
receiver has to check if its current SINR value is compatible with
the $TPR$ rate. Table~\ref{tab:SINR} shows an example of the minimal
(or target) SINRs for using each PHY rate based for the Atheros
chipset~\cite{atheros}. In case the current SINR value is less than
the target SINR, the multicast receiver has to acknowledge the next
receiving data frame. Then, a ACK collision will occur with the
leader, and the AP will use the $Clear Channel Assessment (CCA)$
function to realize that the PHY rate increase is incompatible with
some of the receivers in the multicast group.

\begin{table}[htb]
\begin{center}
\vspace*{0.2cm} \caption{target SINRs for using $TPR$ in IEEE
802.11a.}
\begin{tabular}{|c|c|} \hline
TPR             &  target SINR (dB) \\ \hline \hline 54 Mbps     &
24.56     \\ \hline 48 Mbps     &  24.05     \\ \hline 36 Mbps     &
18.80     \\ \hline 24 Mbps     &  17.04     \\ \hline 18 Mbps     &
10.79     \\ \hline 12 Mbps     &   9.03     \\ \hline $~$9 Mbps
&   7.78     \\ \hline $~$6 Mbps       &   6.02     \\ \hline
\end{tabular}
\label{tab:SINR}
\end{center}
\end{table}

\subsubsection{PHY Rate Adaptation Mechanism}
\label{ssec:adaptation}

\begin{figure}
\begin{center}
\includegraphics[width=8.0cm]{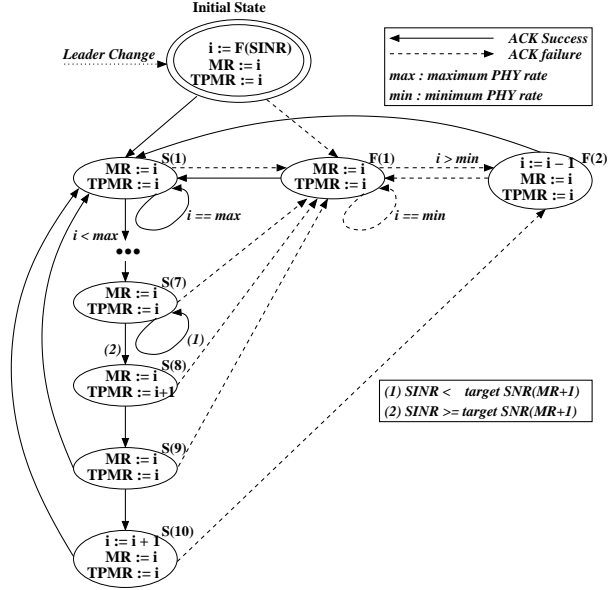}
\caption{State transition diagram for multicast transmission rate
adaptation. } \label{fig:probe}
\end{center}
\end{figure}

The PHY rate adaptation mechanism of RRAM utilizes the state machine
shown in Figure \ref{fig:probe}. In this Figure, solid lines
represent the successful transmissions, dashed lines represent
failed transmissions; $min$ and $max$ stand for the minimum PHY rate
and the maximum PHY rate, respectively. Remark that this state
machine is implemented in the MAC layer of the AP.

\begin{itemize}
\item{\bf Initial State}

In the initial state, the AP chooses the multicast PHY rate
according to the SINR of the new leader. Function $F(SINR)$ returns
the highest PHY rate satisfying the following condition. The SINR
should be larger than target SINR for using the choosed PHY rate.

\item{\bf Success States}

In case of successful transmission, the AP counts the number of
consecutive successful transmission. $S(i)$ stands for the $i^{th}$
consecutive successful transmissions.

However, after the state $S(7)$, the following state is decided
according to the SINR of the ACK frame received in the AP. If the
received SINR of this ACK frame is higher than the target SINR for
using the next higher PHY rate, the state is kept unchanged, $S(8)$.
Otherwise, the state remains at the current state, $S(7)$.

In the state $S(8)$, the AP probes the channel conditions of
non-leader stations to increase the multicast PHY rate. This channel
probe operation consists of two phases.

First, the AP increases the TPR value to the next higher PHY rate as
shown in $S(8)$.

Second, the AP transmits the multicast frame and waits for the ACK
frame. This corresponds to the state $S(9)$. Other stations except
the $Leader$ compare the SINR with the target SINR for using the
TPR. If the SINR of some stations is less than the target SINR,
these stations also temporally become $Leaders$ for this multicast
transmission. Consequently, a ACK collision from several leaders can
occur.

So, if the AP correctly receives the ACK frame, it means that the
SINR of other stations are larger than the target SINR for using
next higher PHY rate.

In the state $S(10)$, the AP increases both the multicast PHY rate
and TPR for probing the channel conditions of the leader. It is
similar to the legacy ARF mechanism because the PHY rate in ARF is
increased after 10 consecutive successful transmissions.

Finally, in the state $S(1)$, if the multicast PHY rate reaches the
maximum PHY rate, the state is kept unchanged.

\item{\bf Failure States}

In case of the transmission failures, the AP also counts the number
of consecutive transmission failures. $F(1)$ and $F(2)$ stand for
$1$ and $2$ consecutive failed transmissions, except for the state
$S(10)$. In the state $S(10)$, once the multicast transmission is
failed, it immediately triggers a PHY rate decrease procedure.

As with ARF, two consecutive failed multicast transmissions triggers
a PHY rate decrease procedure. Especially, in the state $S(10)$, if
the next multicast transmission fails, the state is changed to
$F(2)$ and the multicast PHY rate is immediately decreased. Finally,
in the state $F(1)$, if the multicast PHY rate reaches the minimum
PHY rate, the state is kept unchanged.
\end{itemize}

\subsubsection{Implementation Issues}

First, stations need to know whether the currently associated AP
supports the RRAM mechanism or not. On this purpose, each multicast
receiver checks the duration field of the multicast frame. In the
IEEE 802.11 MAC protocol, the duration field of the multicast frame
is set to 0. If the duration field for the multicast frame is not
equal to 0, this means that the AP supports RRAM.

Second, to implement RRAM, leader stations should turn on the
acknowledgement function for multicast frames. But, most of IEEE
802.11 network interface card do not allow to send ACK frames for
multicast frames. However, such an option could be easily integrated
into the upcoming IEEE 802.11n standard \cite{802.11n}.

\section{Performance Evaluation}
\label{sec:performance}

We evaluate the performance of LB-ARF and RRAM with an extended
version of the NS-2 Simulator \cite{ns-2}\footnote{Simulation codes
and scripts are available at the following URL:
http://www-sop.inria.fr/planete/software/.}. First, in Section
\ref{ssec:static}, we study the $reliability~problem$ and the
$congestion~control~problem$ of the multicast transmission as the
number of multicast receivers increases, with a static environment
scenario. Then, we compare the physical data rate selection
mechanism of legacy IEEE 802.11a, LB-ARF and RRAM, with a static
environment scenario. Second, in Section \ref{ssec:mobility}, we
compare the performance of legacy IEEE 802.11a, LB-ARF and RRAM, in
a mobile environment. Especially, we analyze the multicast
throughput, packet loss rate and multicast rate as the maximum speed
of receivers increases. Futhermore to study the scalability of each
mechanism, we carry out the simulations with a varying number of
multicast receivers, from 5 stations to 25 stations.

Each mobile station is operated in IEEE 802.11a infrastructure mode.
In order to support IEEE 802.11a protocol, we use the enhanced IEEE
802.11a NS-2 module \cite{ns-inria} that comprises the following new
features,

(a) \emph{BER-based~PHY~layer~model} : In the PHY layer, the packet
error rate is determined by the BER and the frame length. In order
to compute the BER, the PHY layer model records SINR variations
during a frame reception. After receiving the frame, the PHY layer
model can compute a more exact BER value even though it requires a
high computation complexity because the BER is recomputed whenever
the SINR is changed.

(b) \emph{IEEE 802.11a multi-rate} : IEEE 802.11a supports 8
different physical data rates, 6Mbps, 9Mbps, 12Mbps, 18Mbps, 24Mbps,
36Mbps, 48Mbps and 54Mbps.

(d) \emph{ARF and AARF \cite{aarf} rate adaptation mechanisms} : ARF
is a well-known rate adaptation mechanism for point-to-point
connection. Adaptive Auto Rate Fallback (AARF) is an extended
mechanism of ARF that improves upon ARF to provide both short-term
and long-term adaptation.

To simulate indoor office environments, we use a log-distance
path-loss model with the path-loss exponent of three \cite{book}.
Additionally, in order to consider multipath fading effect, we use
the Ricean propagation model~\cite{ricean}. When there is a dominant
stationary signal component present, such as with line-of-sight
propagation path, the small-scale fading envelope has a Ricean
distribution.

\subsection{Static Scenarios}
\label{ssec:static}

\subsubsection{Scalability Issues}

A group is composed of five multicast receivers. The number of
unicast stations increases from 1 to 20 stations. All stations are
located near the AP (i.e., 10 m). But, we turn off the rate
adaptation mechanism of LB-ARF and RRAM, so the multicast
transmission rate of each mechanism is fixed to 6Mbps.

\begin{figure}[htp]
\begin{center}
\begin{tabular}{c@{\quad}c}
\vspace*{0.2cm} \hspace*{-0.5cm}
\mbox{\includegraphics[width=4.4cm]{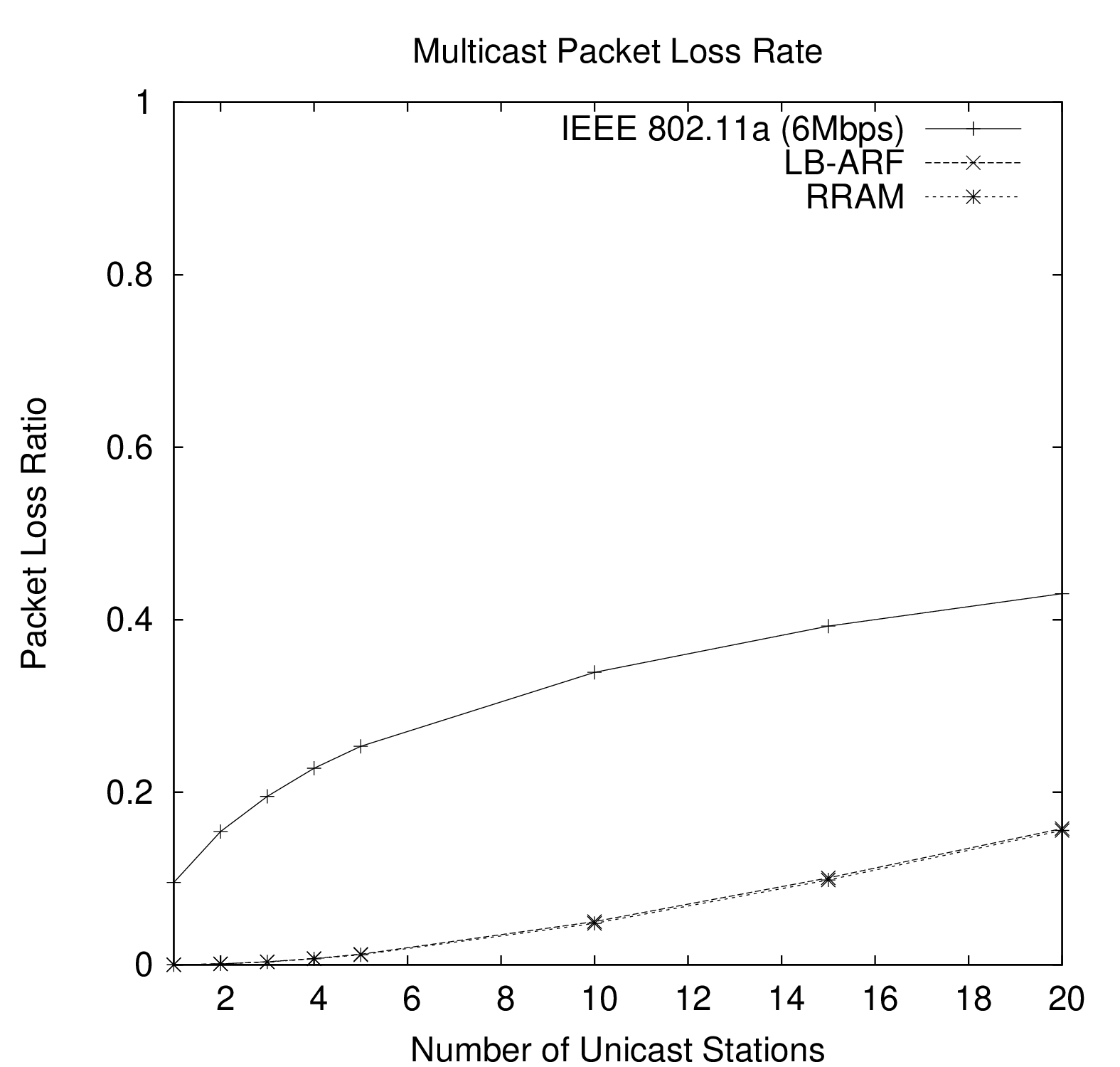}} &
\hspace*{-0.5cm}
\mbox{\includegraphics[width=4.4cm]{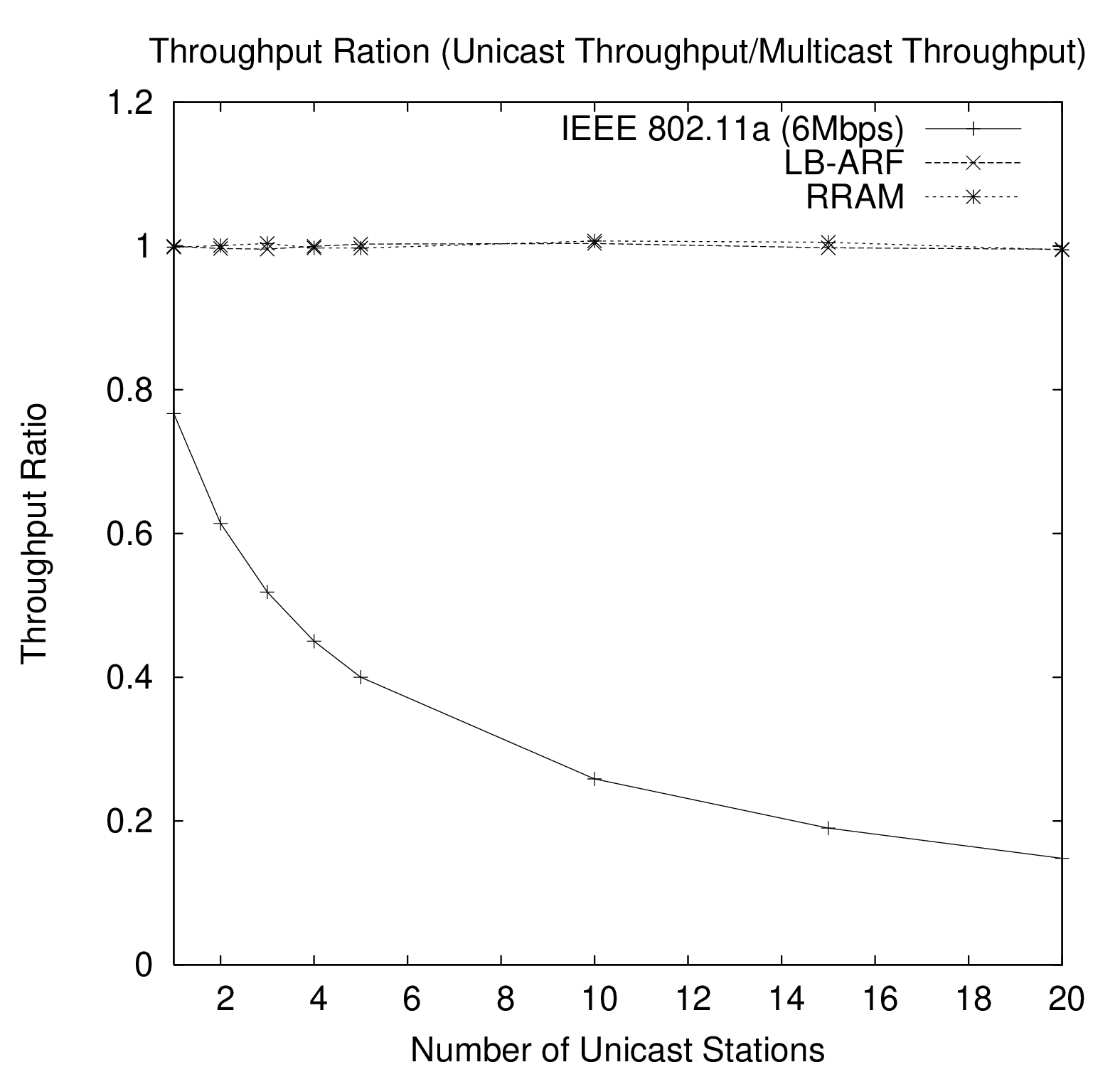}} \\

\small{(a) Packet Loss Rate} &
\small{(b) Throughput Ratio} \\

\end{tabular}
\end{center}
\caption{Fairness performance according to the number of unicast
stations.} \label{fairness}
\end{figure}

We first compute the multicast packet loss rate of each mechanism,
IEEE 802.11a, LB-ARF and RRAM, as the number of unicast stations
increases. As shown in Figure \ref{fairness} (a), the multicast
packet loss rate for the IEEE 802.11a standard is too high, even
when only one unicast station competes with the multicast sender.
LB-ARF and RRAM both obtain lower multicast packet loss rate than
the legacy IEEE 802.11a protocol.

In Figure \ref{fairness} (b), we compute the throughput ratio
between the unicast and the multicast connections, the average
unicast throughput / the average multicast throughput. When the
sender use the legacy IEEE 802.11a protocol, it can not carry out
the binary exponential backoff mechanism. So, we can observe severe
unfairness between the unicast throughput and the multicast
throughput. However, by enabling the binary exponential backoff
mechanism for multicast transmissions, LB-ARF and RRAM are both fair
between unicast and multicast flows.

\subsubsection{Wireless Channel Fading Issue}

We compare the performance of the IEEE 802.11a legacy multicast,
LB-ARF and RRAM, according to the distance between the AP and the
multicast receivers. On this purpose, Five stations join the
multicast group. Initially, the multicast receivers are located to
10 meters away from the AP (with an excellent channel quality), and
only one of the 5 stations is located between 0 and 160 meters from
AP.

Two unicast stations generate a saturated background traffic. These
unicast stations are also located near the AP. We measure the
average throughput of the \emph{best multicast receiver} and the
\emph{worst multicast receiver} that have the best channel
conditions and the worst channel conditions respectively. Although
we measure the average throughput of the unicast stations, we do not
show the results because the results are very similar to the
throughput of the best multicast receiver.

\begin{figure}[htp]
\begin{center}
\begin{tabular}{c@{\quad}c}
\vspace*{0.2cm} \hspace*{-0.5cm}
\mbox{\includegraphics[width=4.4cm]{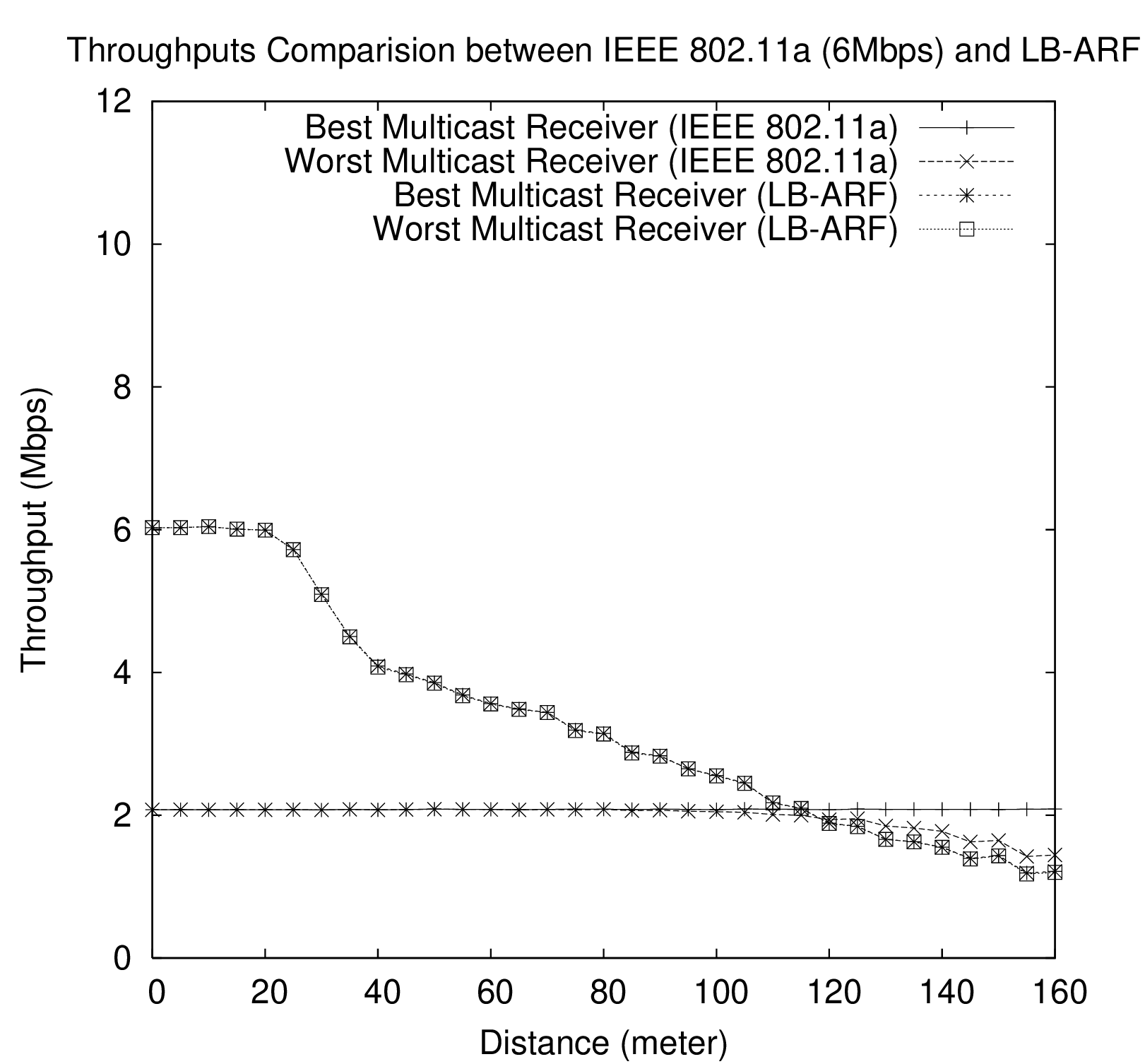}}
& \hspace*{-0.5cm}
\mbox{\includegraphics[width=4.4cm]{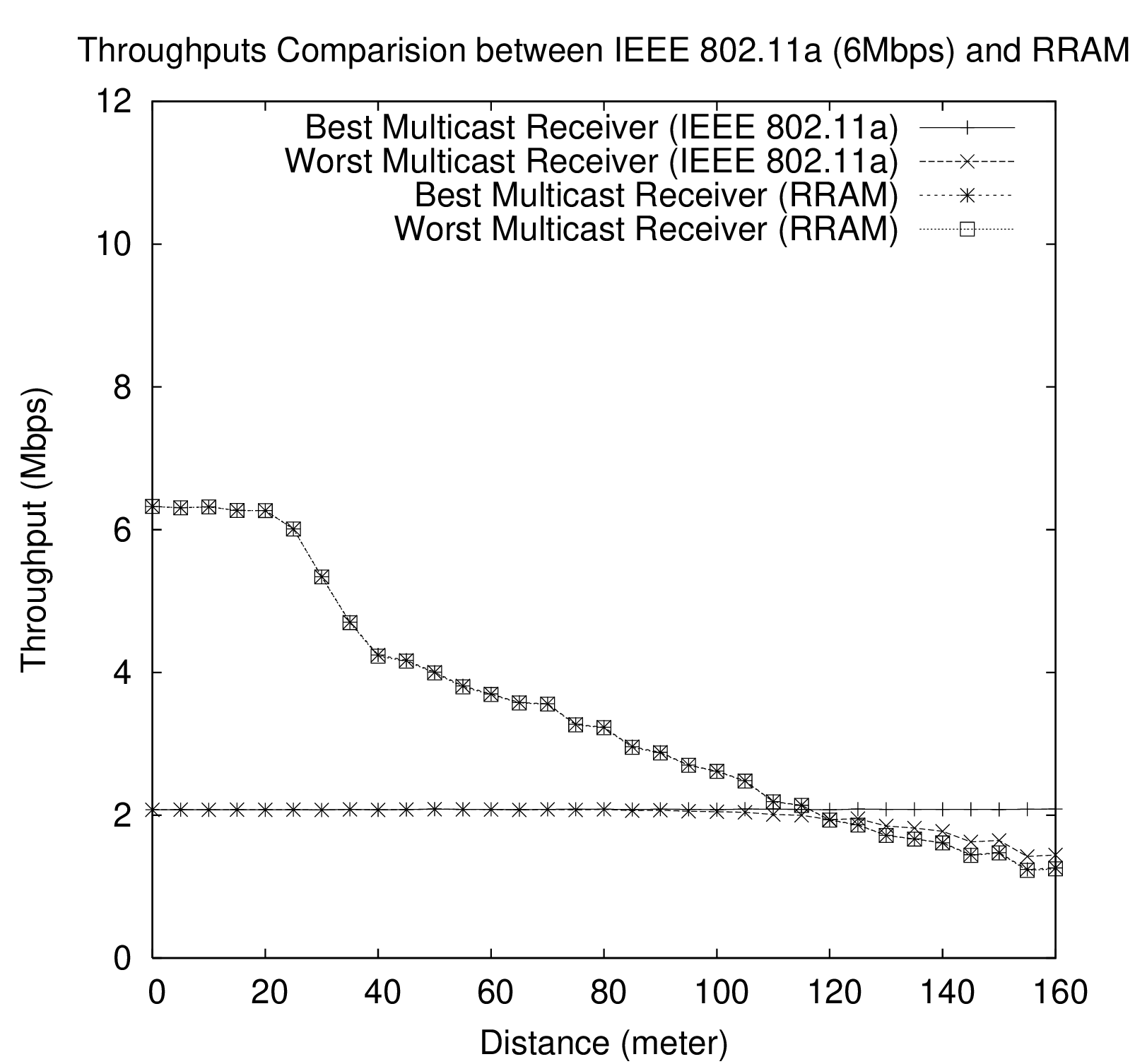}} \\
\small{(a)LB-ARF} & (b) \small{RRAM} \\
\end{tabular}
\end{center}
\caption{Performance comparison of IEEE 802.11, LB-ARF and RRAM with
a Ricean channel model.} \label{simple1}
\end{figure}

We use the Ricean channel model in order to take into account the
multipath fading effect. In Figures \ref{simple1} (a) and (b), we
compare the IEEE 802.11a protocol with both solutions, respectively
LB-ARF and RRAM. The IEEE 802.11a protocol has the lowest throughput
for the multicast connection. When the distance between the worst
receiver and the AP is larger than 100 meters, the worst receiver
experiences some wireless channel errors. Although the multicast
transmission of IEEE 802.11a uses the lowest PHY rate, some frames
are lost because there is no retransmission mechanism for multicast.

In Figure \ref{simple1} (a), when using LB-ARF,  the throughput of
the best multicast receiver and the worst multicast receiver are
similar. It means that LB-ARF correctly chooses the multicast
transmission rate according to the channel conditions of the worst
multicast receiver. In Figure \ref{simple1} (b), when using RRAM,
the worst multicast receiver and the best multicast receiver also
obtain similar throughput than LB-ARF.

\begin{figure}[htp]
\begin{center}
\begin{tabular}{c@{\quad}c}
\vspace*{0.2cm} \hspace*{-0.5cm}
\mbox{\includegraphics[width=4.4cm]{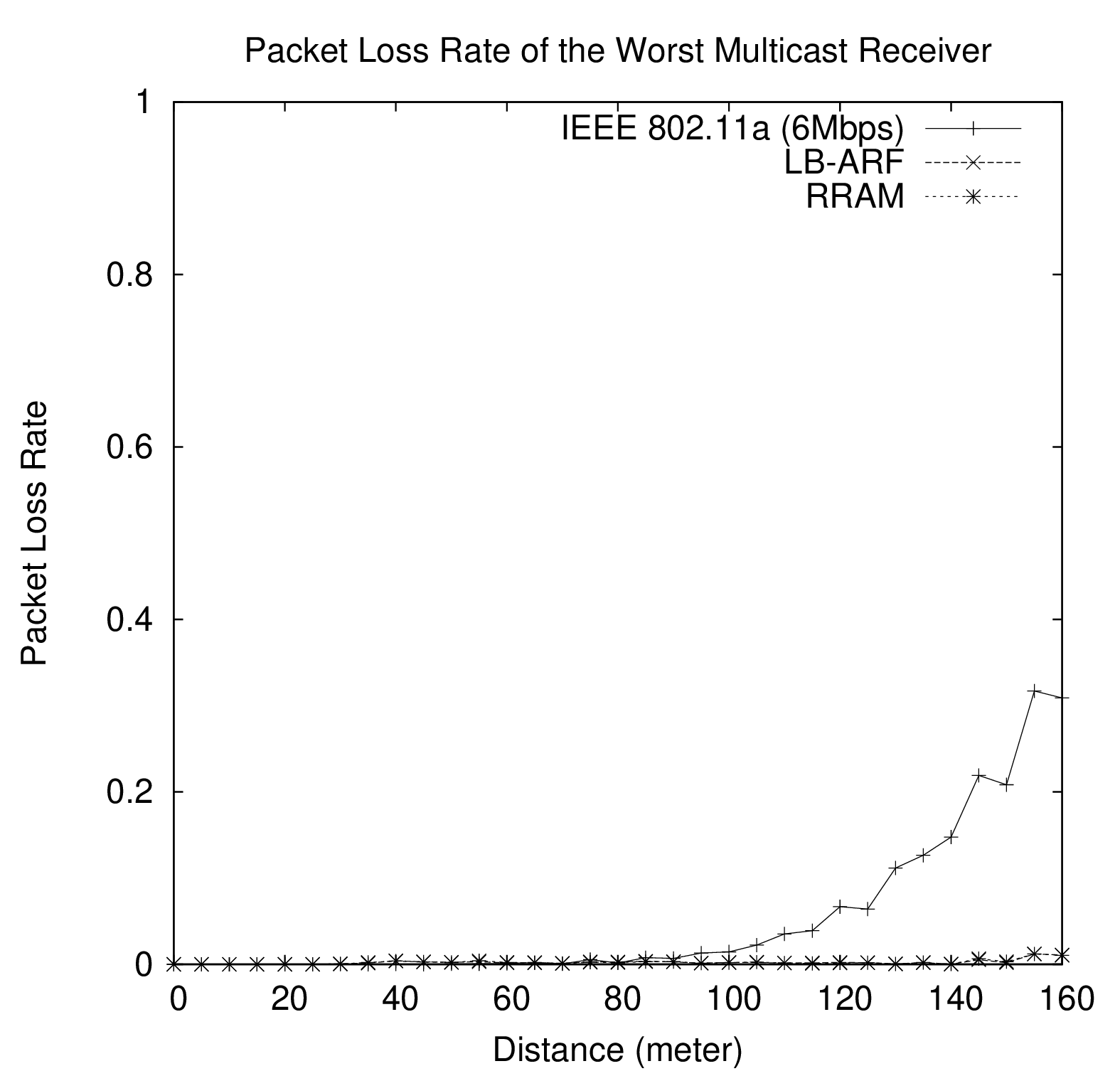}}
& \hspace*{-0.5cm}
\mbox{\includegraphics[width=4.4cm]{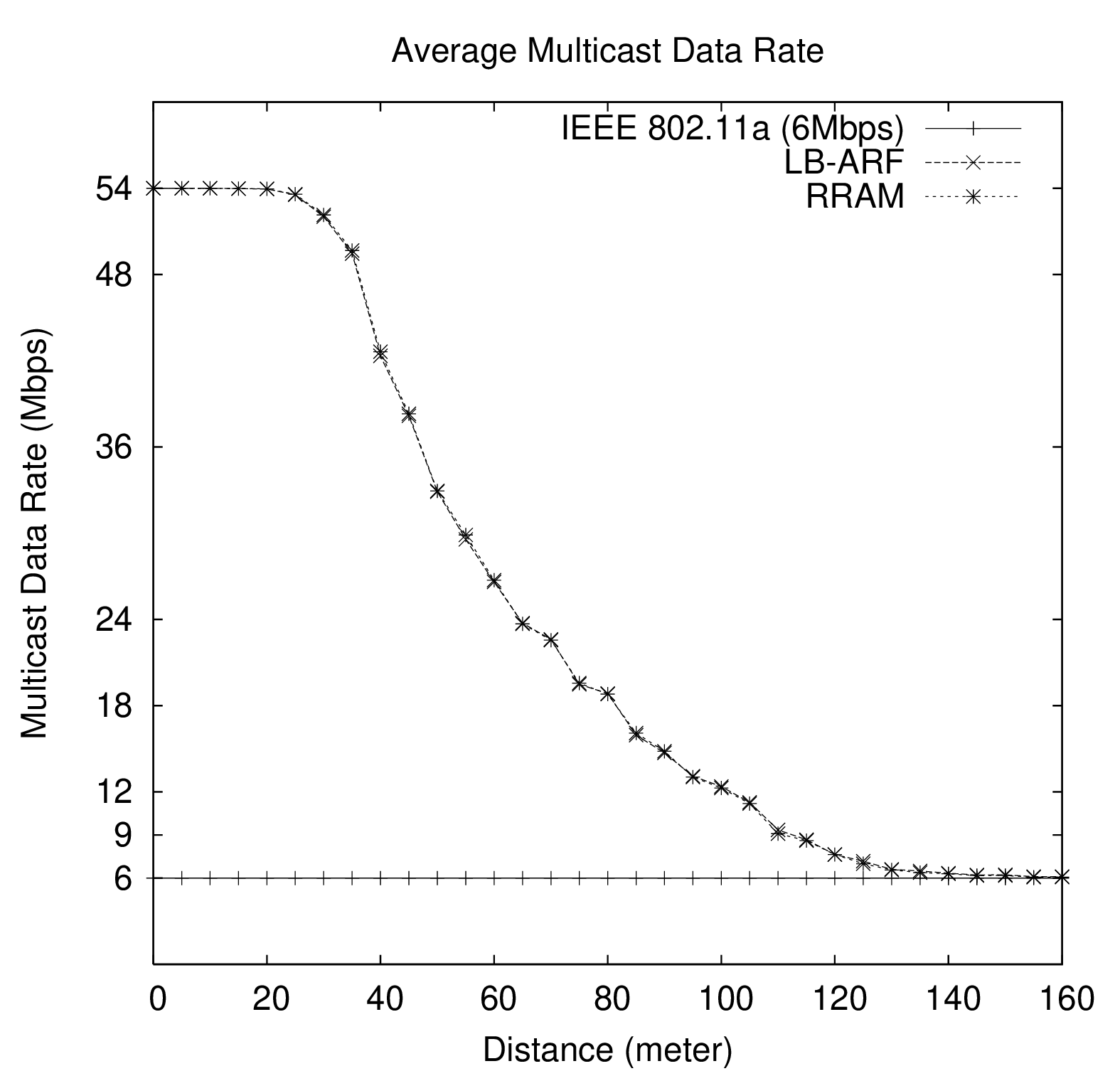}} \\

(a) Packet Loss Rate & \hspace*{-0.7cm}
(b) Average Multicast PHY Rate \\

\end{tabular}
\end{center}
\caption{Packet loss rate and multicast PHY rate comparison.}
\label{simple2}
\end{figure}

Figures \ref{simple2} (a) shows respectively the average packet loss
rates of the worst multicast receiver. The average packet loss rates
of LB-ARF and RRAM are less than 5\%. However, in the case of IEEE
802.11a, the packet loss rate increases after long distance (i.e.,
100 meters), because the IEEE 802.11 MAC protocol does not
retransmit corrupted multicast frames caused by wireless channel
errors. Figures \ref{simple2} (b) shows respectively the average
multicast PHY rates. The average multicast PHY rates of LB-ARF and
RRAM decrease as the channel conditions of the worst station
decreases. Also, there is very few difference between the average
multicast PHY rates of LB-ARF and RRAM. Figures \ref{simple1} and
\ref{simple2} show that LB-ARF outperform the legacy IEEE 802.11a
protocol in static environments, but LB-ARF shows the problem in the
random mobility environements because it is required to more
frquently change leader. We analyze these issues in the following
Section.

\subsection{Mobile Scenarios}
\label{ssec:mobility}

\subsubsection{Scalability Issue}

First we compare the scalability of LB-ARF and RRAM mechanisms by
increasing the number of multicast receivers, with a random mobility
model. The number of multicast receivers increases from 5 receivers
to 25 receivers. One receiver among the set of multicast receivers,
corresponding to the the best receiver, is fixed near the AP. The
others move within a square area (60 m by 60 m) with a random
waypoint mobility model. The maximal velocity of moving receivers is
3 m/s and the pause time is set to 1 s. We also use the Ricean
channel model. Two unicast stations, located near the AP, are used
to generate a saturated traffic.

\begin{figure}[htp]
\begin{center}
\begin{tabular}{c@{\quad}c}
\vspace*{0.2cm} \hspace*{-0.5cm}
\mbox{\includegraphics[width=4.4cm]{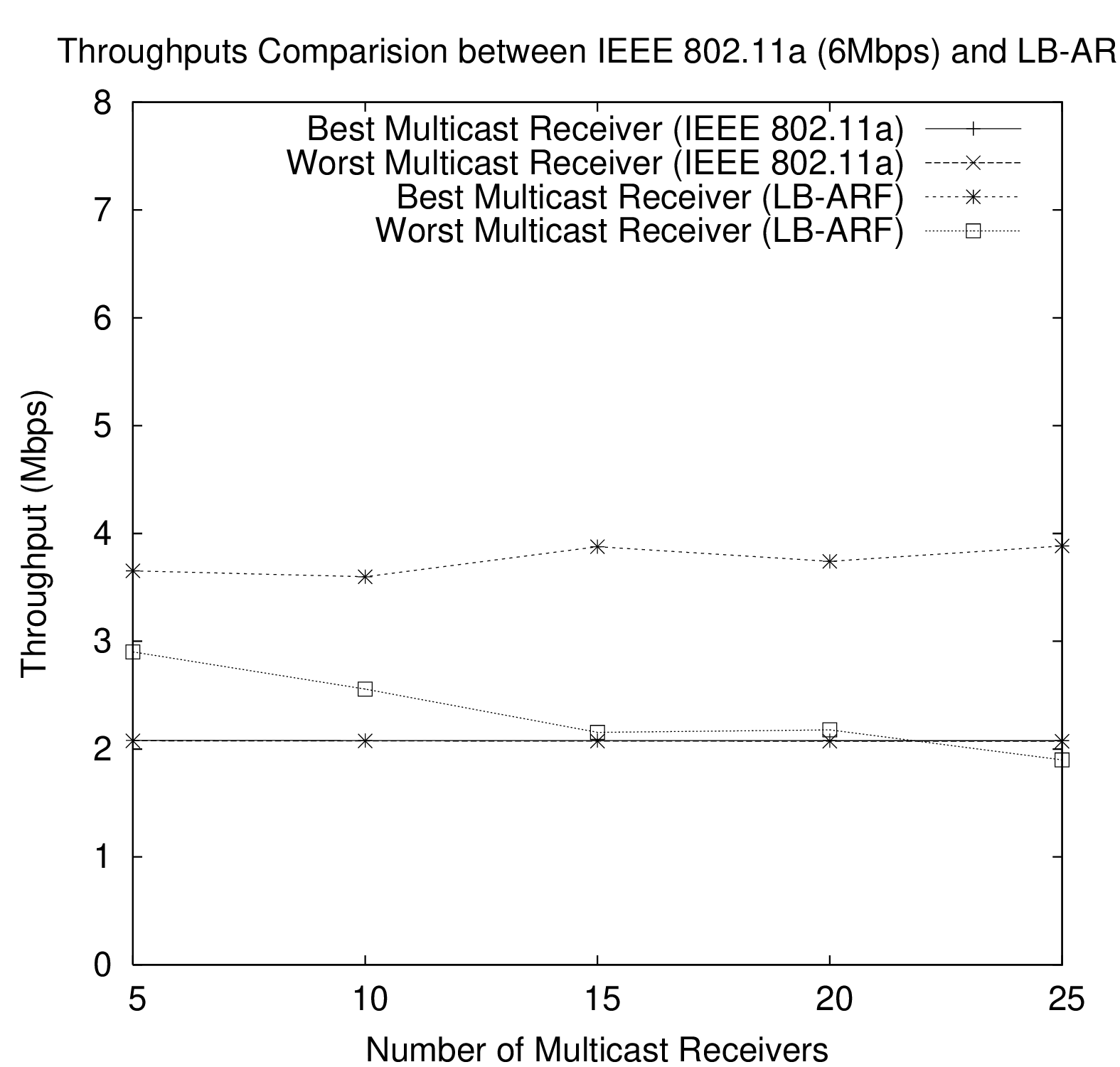}}
& \hspace*{-0.5cm}
\mbox{\includegraphics[width=4.4cm]{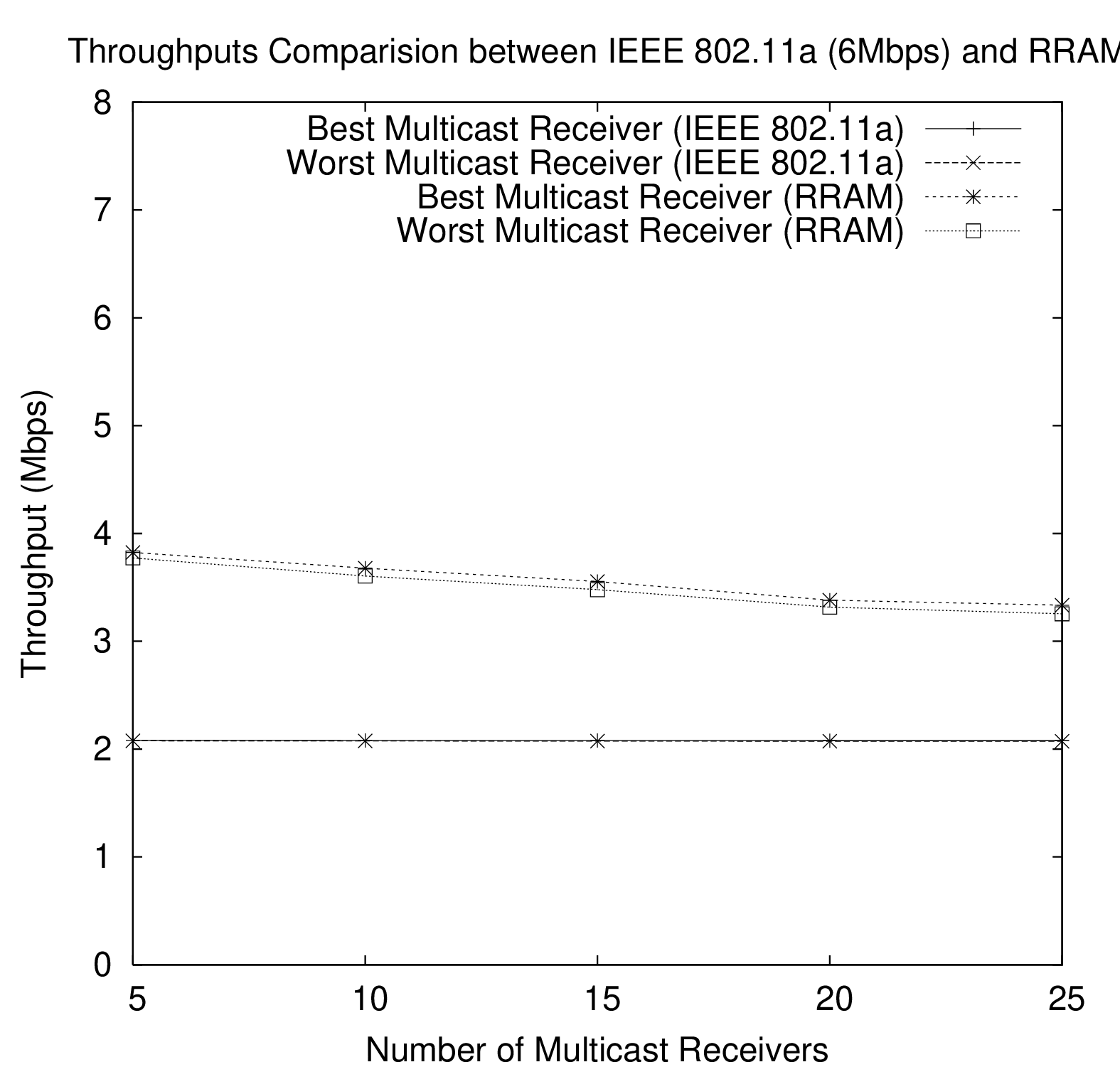}} \\

(a) LB-ARF &
(b) RRAM \\

\end{tabular}
\end{center}
\caption{Performance comparison of IEEE 802.11a, LB-ARF and RRAM.}
\label{random3}
\end{figure}

In Figures \ref{random3} (a) and (b), we compare the IEEE 802.11a
protocol with both solutions, respectively LB-ARF and RRAM. As shown
in Figure \ref{random3} (a), when using LB-ARF, the throughputs of
the best multicast receiver and the worst multicast receiver exhibit
a large difference. Especially, as the number of multicast receivers
increases, the throughput difference between the best multicast
receiver and the worst multicast receivers increases. As the number
of multicast receivers increases, the worst stations is more
frequently changed. But, LB-ARF can not quickly change the leader
whenever the worst multicast receiver changes, because the timescale
of leader election is relatively longer than the one of channel
variation. However, RRAM provides a high transmission reliability
for multicast frames. In Figure \ref{random3} (b), when using RRAM,
the throughputs of the best multicast receiver and the worst
multicast receiver are very similar. Although the number of
multicast receivers increases, RRAM provides the reliabile
transmission and the high throuhgput. It means that RRAM chooses the
appropriate PHY rate for the channel condition of the worst
multicast receiver.

\begin{figure}[htp]
\begin{center}
\begin{tabular}{c@{\quad}c}
\vspace*{0.2cm} \hspace*{-0.5cm}
\mbox{\includegraphics[width=4.4cm]{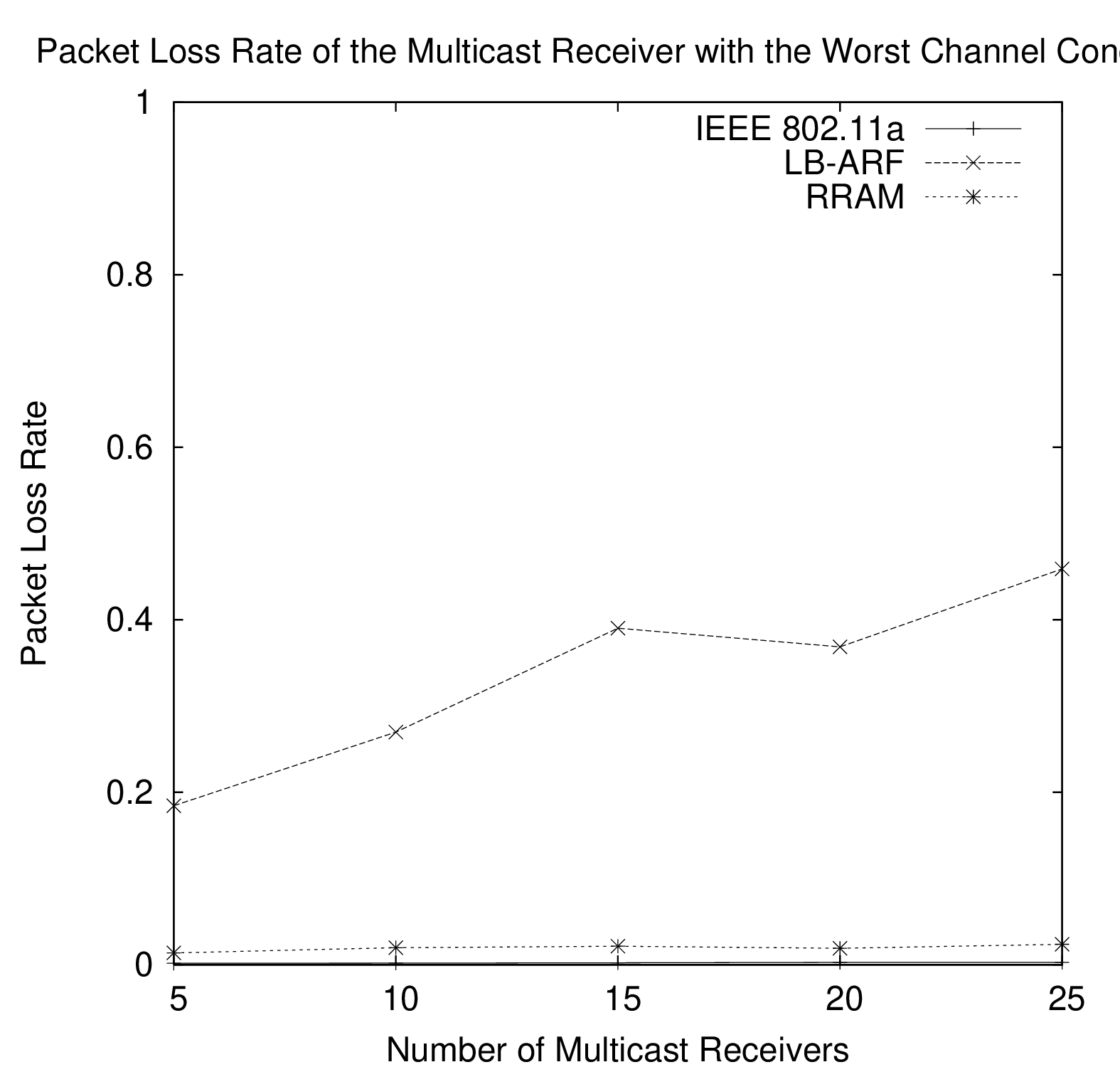}}
& \hspace*{-0.5cm}
\mbox{\includegraphics[width=4.4cm]{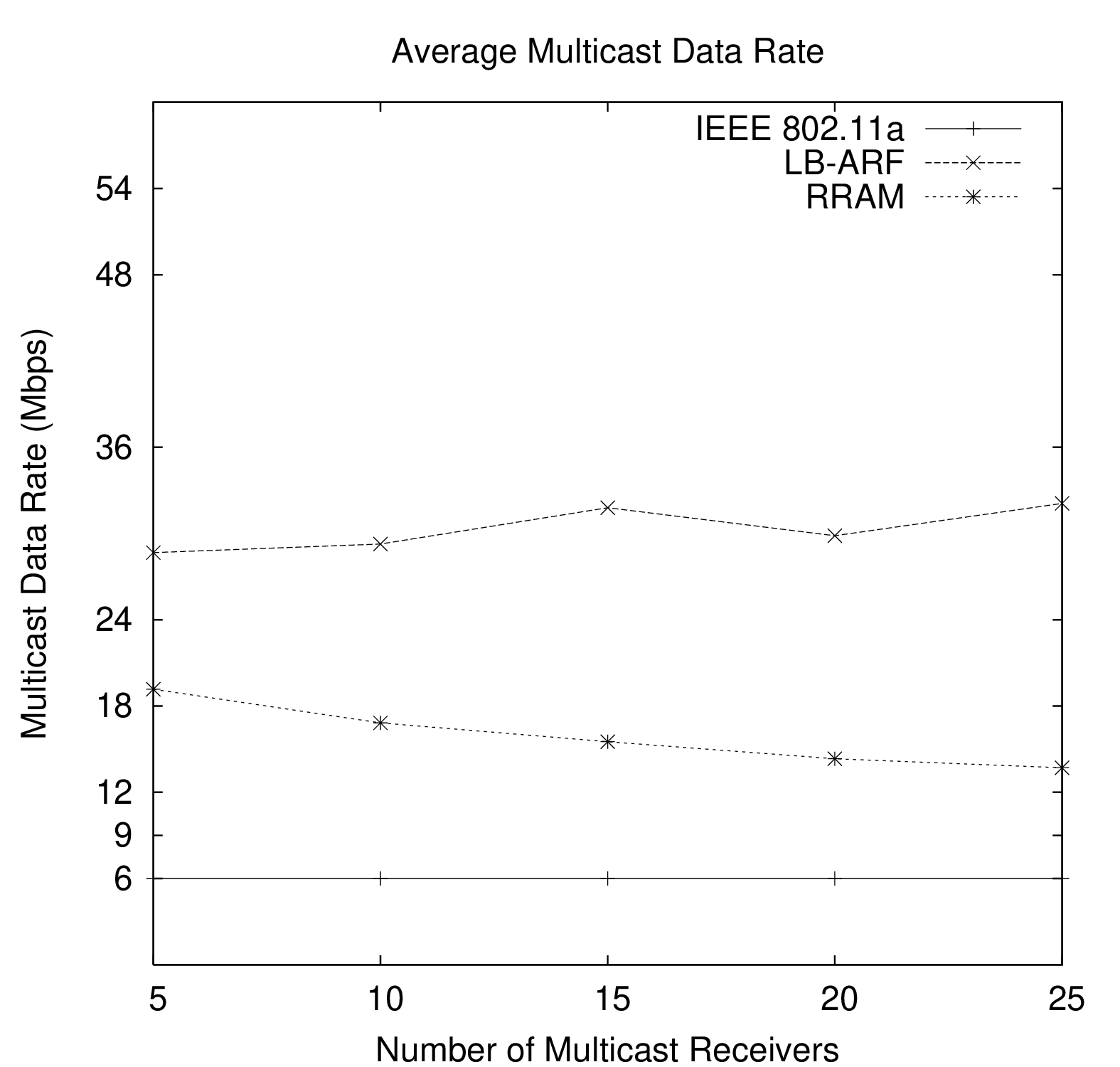}} \\

(a) Packet Loss Rate & \hspace*{-0.7cm}
(b) Average Multicast PHY Rate \\

\end{tabular}
\end{center}
\caption{Packet loss rate and multicast PHY rate comparison.}
\label{random4}
\end{figure}

Figures \ref{random4} (a) shows the average packet loss rates of the
worst multicast receiver. The average packet loss rates of IEEE
802.11a and RRAM are less than 3\%. However, in the case of LB-ARF,
the packet loss rate is larger than 20\% and it also increases as
the number of multicast receivers increases. Because the corrupted
multicast frame of the worst station is never retransmitted, when
the worst stations is not elected to the leader. Figures
\ref{random4} (b) shows the average PHY multicast transmission
rates. In the case of LB-ARF, a high multicast PHY rate is selected.
When the worst station is not selected as the leader, the AP will
use the higher transmission rate. In fact, with LB-ARF, it is
difficult to quickly detect the worst station. However, RRAM allows
to select the appropriate multicast PHY rate that is higher than
6Mbps of IEEE 802.11a. Figures \ref{random3} and \ref{random4},
LB-ARF has the problem in a mobile environment.

\subsubsection{Mobility Issue}

We compare the performance of LB-ARF and RRAM by increasing the
maximum speed of multicast receivers, from 0.1 m/s to 5 m/s. The
number of multicast receivers is fixed to 5 receivers. One of the
multicast receivers is located near the AP. (it corresponds to the
best multicast receiver.) The others move in square area (60 m by 60
m) with a random waypoint mobility model. We also use the Ricean
channel model. Two unicast stations located near to the AP are used
to generate a saturated traffic.

\begin{figure}[htp]
\begin{center}
\begin{tabular}{c@{\quad}c}
\vspace*{0.2cm} \hspace*{-0.5cm}
\mbox{\includegraphics[width=4.4cm]{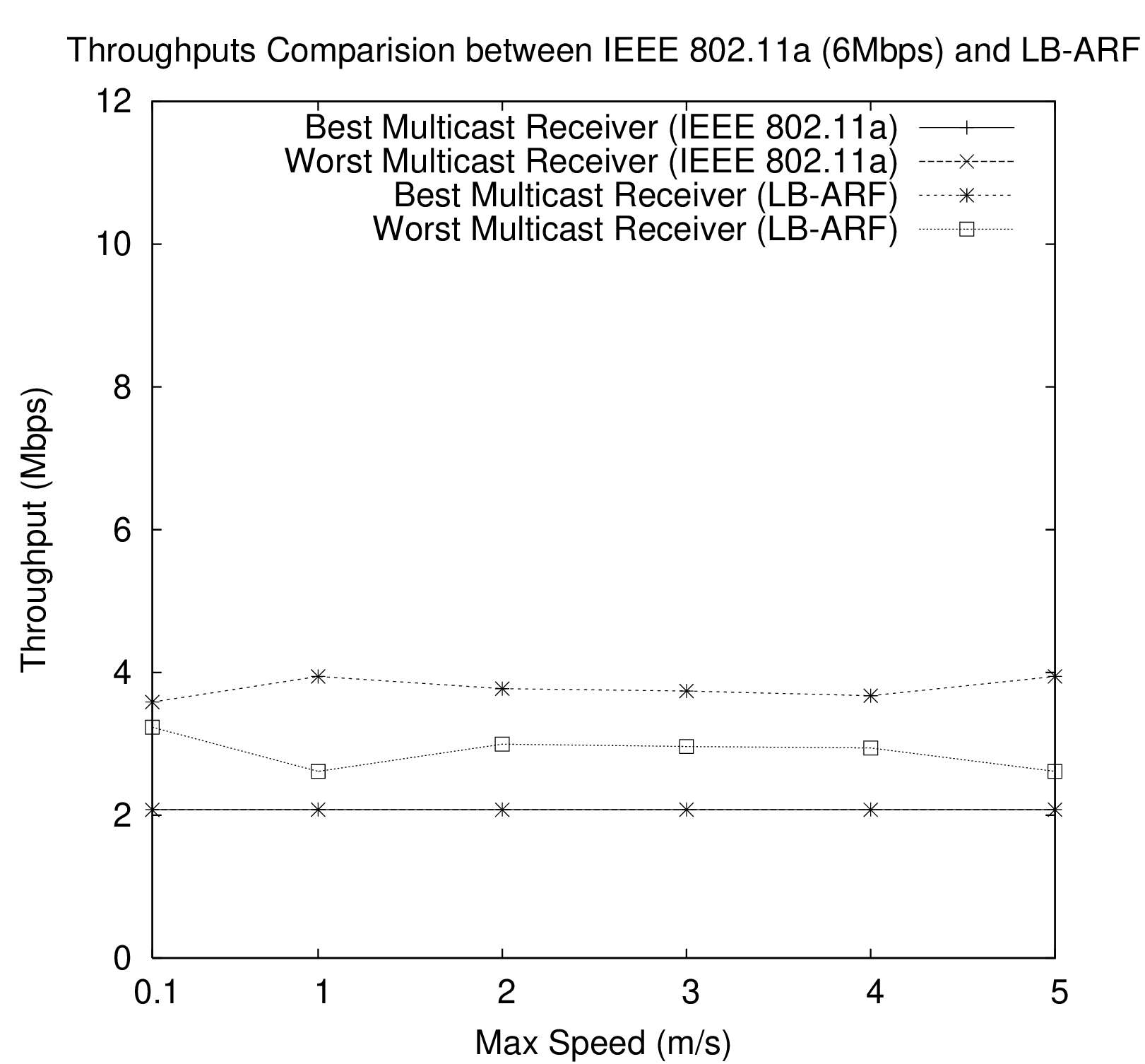}}
& \hspace*{-0.5cm}
\mbox{\includegraphics[width=4.4cm]{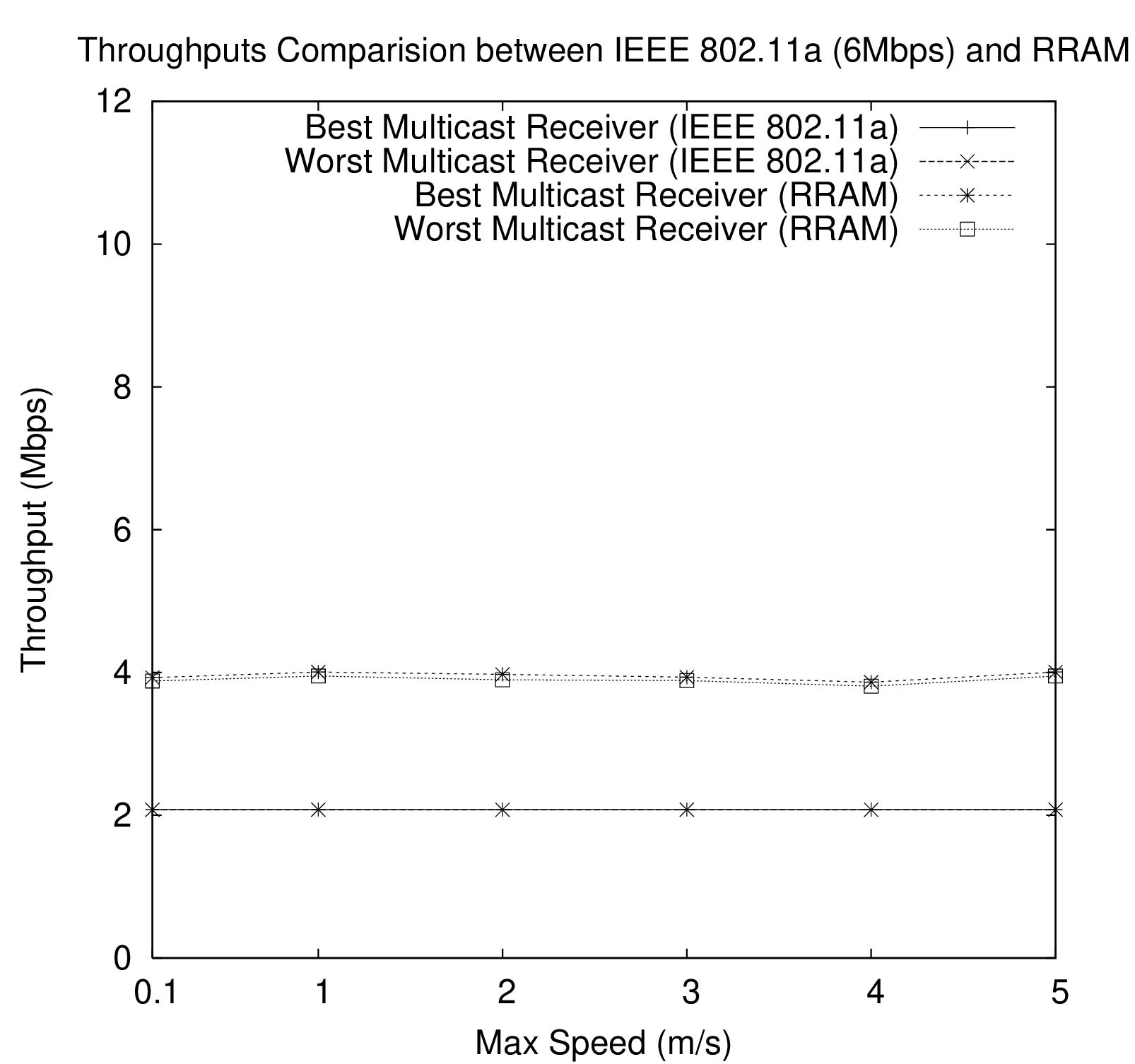}} \\

(a) LB-ARF &
(b) RRAM \\

\end{tabular}
\end{center}
\caption{Performance comparison of IEEE 802.11a, LB-ARF and RRAM.}
\label{random1}
\end{figure}

In Figures \ref{random1} (a) and (b), we compare the IEEE 802.11a
protocol with both solutions, respectively LB-ARF and RRAM. As shown
in Figure \ref{random1} (a), in the case of LB-ARF, the throughput
of the worst multicast receiver is similar than the throughput of
the best multicast receiver at the very low speed such as 0.1 m/s.
However, as the speed increases, we observe a large difference of
throughput between the best multicast receiver and the worst
multicast receiver. Although the number of multicast receivers is
very small, LB-ARF does not achieve throughput fairness between the
set of multicast receivers. In the case of RRAM, as shown in Figure
\ref{random1} (b), the throughputs of the best multicast receiver
and the worst multicast receiver are similar. Especially, even
though the speed of the multicast receivers increases, RRAM provides
high throughput fairness between the best multicast receiver and the
worst multicast receiver. Because RRAM can choose an appropriate
multicast transmission rate even when the worst station is not
elected to the leader.

\begin{figure}[htp]
\begin{center}
\begin{tabular}{c@{\quad}c}
\vspace*{0.2cm} \hspace*{-0.5cm}
\mbox{\includegraphics[width=4.4cm]{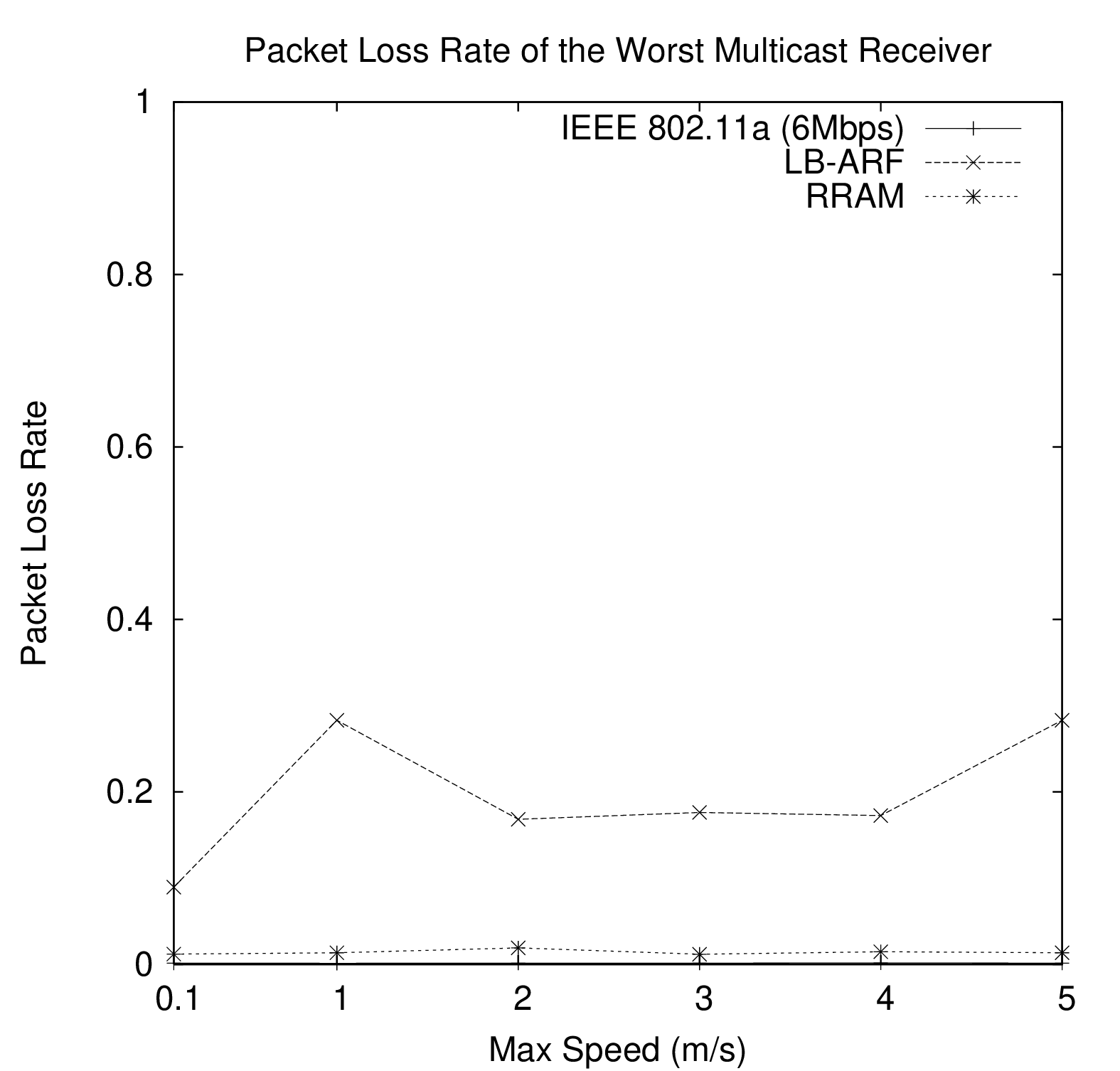}}
& \hspace*{-0.5cm}
\mbox{\includegraphics[width=4.4cm]{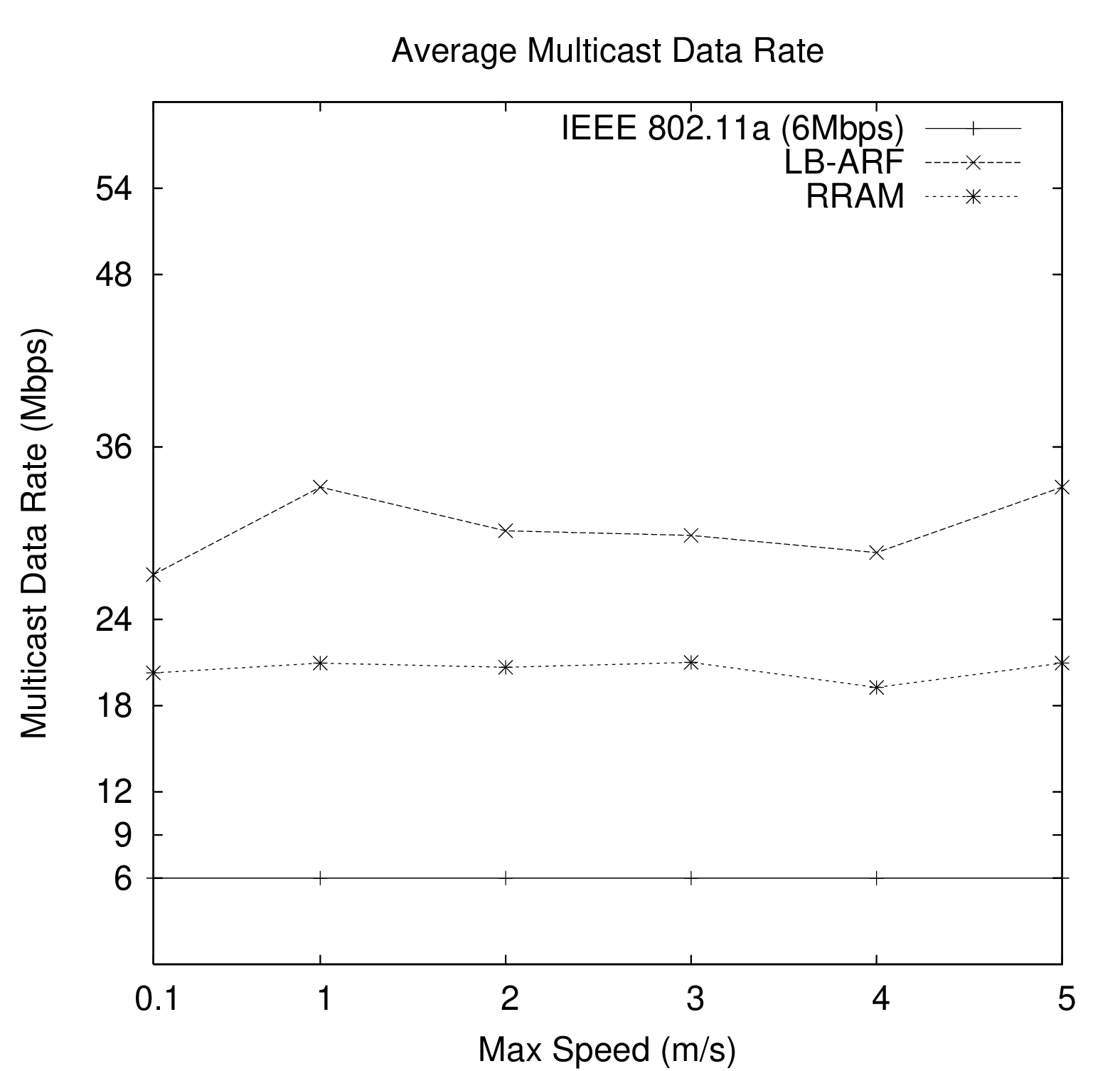}} \\

(a) Packet Loss Rate & \hspace*{-0.7cm}
(b) Average Multicast PHY Rate \\

\end{tabular}
\end{center}
\caption{Packet loss rate and multicast PHY rate comparison.}
\label{random2}
\end{figure}

Figures \ref{random2} (a) and (b) show the average packet loss rates
of the worst multicast receiver and the average multicast
transmission rates, respectively. In the case of LB-ARF, the packet
loss rate of the worst multicast receiver is very high. But, the
packet loss rates of IEEE 802.11a and RRAM are less than 2\%. Even
though the stations' speed increases, RRAM can choose the
transmission rate corresponding to channel conditions of the worst
station. Consequently, the rate adaptation mechanism of RRAM is more
robust even with a large number of multicast receivers or with high
mobility of stations.

\section{Conclusion}
\label{sec:conclusion} In this paper, we propose practical solutions
that make possible the deployment of multicast multimedia
applications in WLANs. Our solutions are based on the leader-based
approach. Because the leader election mechanism cannot run at the
same timescale than the PHY rate adaptation mechanism, we propose
that the PHY rate adaptation mechanism authorizes feedback from any
receivers in the group before taking critical decisions such as rate
increase. We describe and evaluate the LB-ARF mechanism for static
environments and its extended version called RRAM which is efficient
both for fixed and mobile stations.

To implement RRAM, we do not require additional functions such as a
negative acknowledgement in LBP~\cite{lbp} neither new 802.11
control frames such as with ARSM~\cite{jose}. However, our solutions
require the possibility to turn on the acknowledgement function for
multicast frames when necessary.

\section*{Acknowledgments}
This work has been partially supported by the French Ministry of
Research RNRT Project ``DIVINE'' and the Korea Research Foundation
Grant funded by the Korean Goverment (KRF-2005-214-D00340). The
authors wish to thank Katia Obraczka for providing valuable comments
in the paper.

\end{document}